\documentclass[aps,prl,twocolumn,showpacs,preprintnumbers,
amsmath,amssymb,superscriptaddress]{revtex4}

\usepackage{graphicx} \usepackage{color} \usepackage{dcolumn}
\usepackage{bm}

\definecolor{violet}{rgb}{0.5,0,0.5}
\makeatletter \def\@dotsep{5} \makeatother

\newcommand{\tdp}{{\hbox{$\pmb{\pmb{\boldsymbol{\blacktriangledown}}}
 \mkern-20mu$\raise0.15ex\hbox{\textbf{- -}}}}}

\newcommand{\degree}{\ensuremath{^\circ}}

\begin{document}

\title{Electron-mediated ferromagnetism and small spin-orbit interaction in a 
molecular-beam-epitaxy grown n-type $GaAs/Al_{0.3}Ga_{0.7}As$ heterostructure 
with Mn $\delta$-doping}

\author{A. Bove}\affiliation{Physics Department, Duke University, Durham, NC 
27708} \affiliation{Physics Department, Purdue University, West Lafayette, IN 
47906 } 

\author{F. Altomare} \affiliation{NIST, 325 Broadway, Boulder, Colorado 80305} 

\author{N. B. Kundtz} \affiliation{Physics Department, Duke University, Durham, 
NC 27708}

\author{A. M. Chang} \email{yingshe@phy.duke.edu} \affiliation{Physics 
Department, Duke University, Durham, NC 27708}

\author{Y. J. Cho} \author{X. Liu} \author{J. Furdyna} \affiliation{Department 
of Physics, University of Notre Dame, Notre Dame, Indiana 46556} 

\date{\today}

\begin{abstract}
We report on transport measurements indicating electron-mediated ferromagnetism 
in a molecular-beam-epitaxy (MBE) grown $GaAs/Al_{0.3}Ga_{0.7}As$ 
heterostructure with Mn $\delta$-doping. The interaction between the magnetic 
dopants (Mn) and the Two-Dimensional Electron Gas (2DEG) gives rise to magnetic 
ordering at temperatures below the Curie temperature ($T_{C}\sim$1.7K) when the 
2DEG is brought in close proximity to the Mn layer by gating. The Anomalous Hall 
Effect (AHE) contribution to the total Hall resistance is shown to be three 
orders of magnitude smaller than in the case of hole-mediated ferromagnetism 
indicating the near absence of spin-orbit interaction.
\end{abstract}

\pacs{75.47.-m, 75.50.Pp, 85.35.Be} \keywords{DMS, LT-MBE, GaMnAs, Ohmic 
contact}
\maketitle

In diluted magnetic semiconductor (DMS) systems, when a high concentration of 
magnetic impurities is incorporated in the non-magnetic semiconductor hosts, 
ferromagnetic behavior is observed. In these systems, magnetism is produced by a 
carrier-mediated interaction between magnetic impurities. In DMS systems such as 
the promising ferromagnetic GaMnAs systems, the origin of ferromagnetism has 
been proven to be hole-mediated (\cite{OhnoScienceRev}, \cite{Ohno1992}).   

To date, the only possibilities for n-type, electron-mediated ferromagnetism are 
found in the large bandgap GaMnN system (\cite{GaMnN1},\cite{GaMnN2})and the 
ZnMnAlO system \cite{ZnMnAlO}. Even though evidence of ferromagnetism is 
observable at temperatures above room temperature, due to a scarcity of magneto-
transport data, the claim that magnetism is carrier-mediated rather than 
originating from the clustering of the Mn atoms has remained controversial.  

In this letter, we present direct evidence in magneto-transport measurements demonstrating 
electron-mediated ferromagnetism in a specially designed, low carrier 
concentration ($\sim10^{12}cm^{-2}$) GaMnAs quantum well (QW). In our structure, 
the ferromagnetism is controllable via tuning of the n-carrier density.  The 
ferromagnetism is manifested in unambiguous, hysteretic behavior of the 
transport coefficients under the application of a magnetic field.  Hysteresis is 
present only when the GaMnAs quantum well is filled with electrons via gating, 
and is absent when the electrons only occupy a heterojunction (HJ) situated far 
away from the Mn atoms. Moreover, in our system, there is a clear absence of the 
anomalous Hall effect (AHE) -an effect rising from spin-orbit scattering in the 
presence of a finite magnetization.  This absence stands in contrast to what has 
been widely observed in all other GaMnAs systems, which were all p-type systems, 
and is consistent with a GaAs conduction band, which is primarily s-band in 
character.

Our n-type crystal containing a $\delta$-doping Mn layer in a GaAs QW was grown 
by molecular-beam-epitaxy on a semi-insulating (SI) GaAs (001) substrate (see 
table 1 for the growth structure). The 2/3 monolayer (ML) of Mn was grown at the 
GaAs/AlGaAs interface situated closer to the crystal top surface while keeping 
the substrate at low temperature (LT) $\sim$250\degree C. 
\begin{table}[th]
	\centering \caption{Heterostructure}
	\begin{tabular}[t]{ll}
						GaAs (001)& Substrate\\
						GaAs:Si& 100 nm\\
            AlGaAs:Be& 200 nm\\
					  SI-GaAs& 10 nm\\
					  LT-Mn& $\frac{2}{3}$ ML\\
					  LT-AlGaAs& 10 nm\\
					  LT-AlGaAs:Be& 15 nm\\
					  LT-GaAs& 5 nm\\
	\end{tabular}
	\label{tab:magneticstructure}
\end{table}

Based on previous studies (\cite{2DHG}, \cite{2DHG2}), at such low growth 
temperature the Mn atom should be free of clustering, and are uniformly 
distributed. In this structure, a layer of 2D electron gas (2DEG) is always 
present at the interface of the Si doped 200 nm GaAs and the 100 nm AlGaAs:Si/Be 
layers, $\sim$140 nm below the surface, while the 10 nm GaMnAs QW can be filled 
or emptied by gating.

To fabricate devices for transport measurements, it is necessary to make good 
ohmic contact. A novel method was developed for annealing ohmic contact, which 
employs a thermal gradient along the sample. This method is able to preserve the 
ferromagnetism in the center of the sample, where the Hall bar is located, while 
destroying it in the vicinity of the contacts \cite{condmat}. Following 
annealing of the contacts, standard lithographically and wet-etching were 
performed to define a Hall bar with a 150 $\mu$m wide channel. The voltage 
probes were narrow, 10 $\mu$m in width, to minimize their perturbation on the 
current flow. Magneto-transport measurements were carried out using a lock-in 
amplifier at 3.7 Hz and 10 nA excitation current.

\begin{figure}[h]

\centering \includegraphics[clip, width=3.1in]{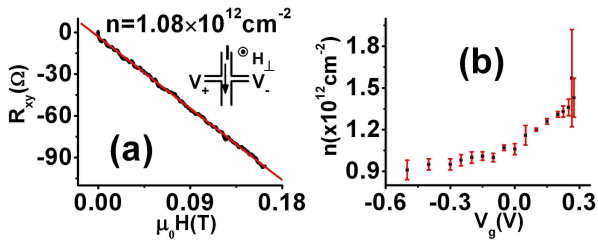}
\centering \caption{(a) Carrier density of the 2DEG at T=4.2K. (b) Gate voltage 
dependence of the carrier density (Note the change in slope near $\sim$0V, 
indicating filling of the GaMnAs QW).  Note that the $R_{xy}$ was anti-
symmetrized with respect to H to remove residual contributions from $R_{xx}$}

\label{fig:fig1}

\end{figure}

The first task is to establish the carrier type. The sign of the Hall resistance 
($R_{xy}$) in Fig. 1(a) indicates that our structure contains n-type carriers.  
An estimate of the carrier density using a one component carrier model to 
analyze the Hall coefficient ($n \propto 1/R_{Hall}$ rather than a two component 
model, one in the QW and the other at the Si-doped GaAs and AlGaAs:Si/Be 
interface) yielded a carrier density $n\sim1.08\times10^{12}cm^{-2}$, and a 
mobility $\mu\sim575cm^{2}/Vs$ at 4.2K \cite{condmat}. This density has a 
sizable error of $\sim50\%$ arising from the assumption of one-component 
\cite{condmat}, but gives a rough characterization when the mobility and density 
in the two components are roughly comparable, as is the case here. To further 
establish that the carrier type inside the QW is the electron, the gating 
dependence of the $R_{xy}$ was examined after deposition of a metal gate.  The 
resultant carrier density is shown in Fig. 1(b) as a function of the applied 
voltage. A more positive gate voltage increases the density, clearly 
establishing the carrier type to be n-type. Notably, there is an abrupt change 
in slope near $V_{g}\sim0$ V, indicating the filling of the GaMnAs quantum well 
(QW) for positive gate voltages. The slopes for $V_{g}>0$ V and $V_{g}<0$ V, 
yield capacitances corresponding to a 2DEG distance of $\sim50\pm10$ nm and 
$\sim210\pm30$ nm, compared to the nominal distances of 35 nm and 145 nm, 
respectively \cite{condmat}. The distances, $40\%$ larger than nominal, could 
result in part from a higher crystal growth rate than nominal, which was 
calibrated to $\sim25\%$ accuracy.

To investigate the magnetism, our device was cooled to 0.3 K and an in-plane 
magnetic field $H_{||}$ was applied. At zero gate voltage, no hysteresis was 
observed in $R_{xy}$ (black traces in Fig. 2(a)). However, after applying a 
positive gate voltage of 500 mV to fill the GaMnAs QW with electrons, a clear 
hysteretic loop appeared in the hall signal (red traces in Fig. 2(b)).  Similar 
behavior was observed in several Hall junctions and in the longitudinal 
resistance $R_{xx}$.  At the same time, $R_{xx}$ exhibited a small but 
reproducible feature as a function of temperature, indicative of a ferromagnetic 
transition near 1.7 K (Fig. 2(c)).
\begin{figure}[h]

\centering \includegraphics[clip, width=3.2in]{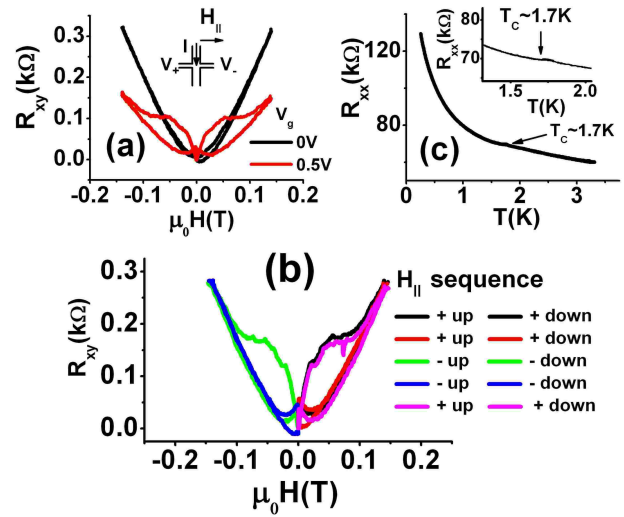}
\centering \caption{
(a) Hysteresis loops (red traces) are observed after a gate voltage of 0.5V is 
applied at T=0.3K. (b) History dependent behavior at T=0.3K with an in-plane 
magnetic field applied. When magnetism appears, the $R_{xy}$ traces often no 
longer appear anti-symmetric in B, despite the well-defined Hall bar geometry.  
(c) Longitudinal resistance as a function of the temperature. The bump at around 
1.7K is attributed to the ferromagnetic phase transition.}

\label{fig:fig2}
\end{figure}

To further substantiate the hysteretic behavior, the magnetic field was swept up 
and down twice in the positive direction before changing its sign. When the 
magnetic field was first swept up and down (black traces), a clear loop 
appeared, while for the traces taken subsequently without reversing the magnetic 
field (red traces), the loop is absent. Similar behavior was observed for 
negative field traces (green and blue, respectively). 

To elucidate the n-type mediated ferromagnetism, we studied the AHE and compared 
it with reported systems where ferromagnetism is known to be hole-mediated. To 
date, experimental studies of the AHE in electron-mediated systems are still 
lacking, (\cite{GaMnN1}, \cite {GaMnN2} and \cite{ZnMnAlO}), while there are 
only a few theoretical predictions 
(\cite{JungwirthPRL2002},\cite{JungwirthRMP}). As a result of a much reduced 
spin-orbit coupling, it is believed that the AHE should either be absent or much 
smaller than in hole-mediated systems \cite{JungwirthPRL2002}.

For this purpose, a new device was prepared and measured the in a perpendicular 
field $H_{\perp}$ versus temperature. The advantage of this second device 
resides in the fact that possibly due to either inhomogeneity in the local 
strain from the LT growth or a gradient in the Mn concentration, this second 
sample had areas that showed ferromagnetism while others did not as it can be 
seen in Figs. 3(a) and 3(b).  Note that without strain or residual dipole 
interaction to introduce anisotropy in the Heisenberg spin coupling between Mn 
impurities, there should, in principle, be no ferromagnetic state at a finite 
temperature for this 2D system.\cite{DasSarma}  We are thus able to study the 
AHE for a purely paramagnetic (PM) system (non-magnetic junction) and for 
another that shows a ferromagnetism (FM) (magnetic junction) when the 
temperature is below the Curie temperature $\sim 2K$ (Fig. 4(a)).

\begin{figure}[h]

\centering \includegraphics[clip, width=3.1in]{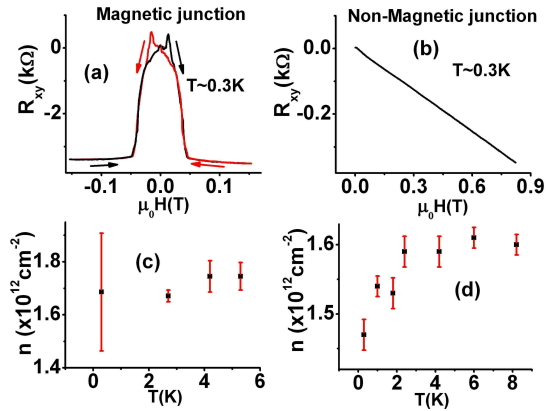}
\centering \caption{(a) Hall resistance for the ferromagnetic junction. (b) Hall 
resistance for the paramagnetic junction. (c) Carrier's density for the 
ferromagnetic junction as a function of temperature. (d) Carrier's density for 
the paramagnetic junction as a function of temperature. The magnetic field is 
perpendicular to the sample's surface in all four cases.}

\label{fig:fig3}

\end{figure}

The Hall resistance $R_{xy}$ versus $H_{\perp}$ was measured at different 
temperatures for both the FM and PM junction and the density deduced from $n 
\propto 1/R_{Hall}$ are plotted in Figs. 3(c) and 3(d), respectively. A detailed 
two-component analysis of the Hall coefficient $R_{Hall}$ is presented below.  
For the FM junction, the density (Fig. 3(c)) remained constant within 
experimental error. On the other hand, for the PM junction the density shown in 
Fig. 3(d) remained constant above 2 K, but decreased as the temperature was 
further reduced down to $T\sim0.3K$. This latter behavior is similar to what is 
expected from an AHE contribution to the Hall resistivity 
(\cite{Ohno1992},\cite{JungwirthRMP}). Jungwirth et al. \cite{JungwirthRMP} have 
suggested that a true AHE should saturate at some high field value and the real 
Hall coefficient should be restored above such a field. To distinguish an AHE 
contribution from the magnetic freeze-out of carrier, we examined the Hall 
resistivity at higher magnetic fields, beyond the saturation of Mn spins.  
Lowering the temperature to $T\sim0.3K$ and sweeping the magnetic field up to 
$\sim$0.8 T, the $R_{xy}$ did not show any appreciable non-linearity in that 
field range, indicating that the AHE contribution must be small.

To quantify the AHE coefficient, we fitted the $R_{Hall}$ as a function of 
temperature for the PM junction using the standard expression for the Hall 
resistivity $\rho_{Hall}$: 
\begin{equation}
	\rho_{Hall}= R_{Hall} \mu_o H = R_{0}\mu_o H +R_{s}M
\end{equation}
where $R_{0}$ is the Ordinary Hall Effect (OHE) coefficient, $R_{S}$ is the AHE 
coefficient and M is the magnetization. M for a paramagnet is proportional to 
the magnetic susceptibility $\chi$ that obeys the Curie-Weiss law $\chi=C/(T-
T_{C})$. C is the Curie constant determined in \cite{FurdynaJAP} and is 
theoretically expressed as $C=(4p^{2}_{eff}\mu^{2}_{B}x/3a^{3}_{0}k_{B})$ where 
$x$ is the Mn concentration, which in our case is $x\sim33.3\%$. For the PM 
junction we set $T_{C}\sim0K$.
\begin{figure}[h]

\centering \includegraphics[clip, width=3.1in]{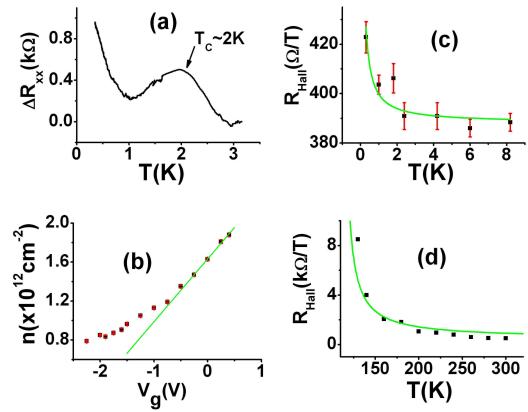}
\centering \caption{(a) Longitudinal resistance as a function of temperature 
gives a Curie temperature value of about 2K. (b) Carrier's density as a function 
of temperature for a ferromagnetic junction. AHE fitting using (c) 
$R_{Hall}=R_{0}+C/\mu_{0}T$ in our sample and (d) 
$R_{Hall}=450\Omega/T+C/\mu_{0}(T-112K)$ in the sample by \cite{2DHG}.}

\label{fig:fig4}

\end{figure}

The fit shown in figure 4(c) yielded $R_{0}=388\Omega/T$, and 
$R_{S}\sim2.73\pm0.58\times10^{-11}\Omega/T$ for the AHE coefficient. Comparing 
to the AHE coefficient $R_{S}\sim4.46\pm0.76\times10^{-7}\Omega/T$ for a p-type 
GaMnAs system, extrapolated from the data of Nazmul et al. \cite{2DHG} (figure 
4(d)), we find our $R_{S}$ value to be $\sim$16,000 times smaller. To obtain a 
more accurate bound, it is necessary to account for the two conducting 
components in parallel. Since the AHE coefficient is associated only with the 
electrons residing in the QW, it is their contribution that needs to be 
evaluated. We begin by expressing the Hall coefficient of the two components in 
the following form:
\begin{equation}
R_{Hall}=-
\frac{n_{QW}\mu^{2}_{QW}+n_{HJ}\mu^{2}_{HJ}}{e\left(n_{QW}\mu_{QW}+n_{HJ}\mu_{HJ
}\right)^{2}}
\end{equation}
where e is the magnitude of the electron charge, $n_{QW}$ the electron density 
in the QW, $n_{HJ}$ the electron density in the HJ and, and $\mu_{QW}$ and 
$\mu_{HJ}$ are the respective mobilities. In this sample, a one-component 
analysis of the density dependence on gate voltage near $V_g \sim 0 V$ signified 
a depth for the 2DEG that is in between the depths of the QW and the HJ. (See 
Fig. 4(b)), indicating that both two channels are contributing. To disentangle 
the two contributions we note that Eq. 2 indicates that the fractional size of 
the QW contribution to $R_{Hall}$ is 
$\left(n_{QW}\mu_{QW}^2\right)/\left[n_{QW}\mu_{QW}^2+n_{HJ}\mu_{HJ}^2\right]=qr
^2/\left(qr^2+1\right)$, where $q=n_{QW}/n_{HJ}$, and $r=\mu_{QW}/\mu_{HJ}$. 
From Fig. 4(b) the slope for the density versus $V_{g}$ at $V_{g}=0$ V, in a 1-
component analysis, i.e. $n\propto\left(1/R_{Hall}\right)$, is 
$0.75\pm0.1\times10^{12}cm^{-2}/V$.  This compares to an expected slope of 
$1.44\times10^{12}cm^{-2}/V$ for the QW 2DEG at a depth of 50 nm below the 
surface.  This reduced slope by a factor of 1.9 implies a depth of $\sim96$ nm, 
which is at 1.9 times the depth of the QW 2DEG but $\sim 1/2$ times that for the 
HJ 2DEG. As $V_{g}$ is reduced below 0 V, down to $V_{g}=-1.5$ V, the slope 
gradually decreases before reaching the value expected for the HJ as the QW 
becomes empty. This difference of 1.5 V indicates that at $V_{g}=0$ V, there is 
a density of $\sim2.15\times10^{12}cm^{-2}$ in the QW. The HJ density at 
$V_{g}\ge-1.5$ V is $0.85\times10^{12}cm^{-2}$. At $V_g = 0 V$, this gives a 
ratio between the two densities of $q=n_{QW}/n_{HJ}\sim2.15/0.85=2.53$. Since 
$n_{QW}$ is linearly proportional to the change in $V_{g}$ we take the 
derivative of $1/R_{Hall}$ with respect to $n_{QW}$ in Eq. (2) to compare to the 
experimental slope:
\begin{equation} 
\frac{d\left(1/R_{Hall}\right)}{dn_{QW}}=e\frac{\left[q^{2}r^{4}+2qr^{2}+2r-
r^{2}\right]}{\left(qr^{2}+1\right)^{2}} .
\end{equation}
Setting it to e/1.9 at $V_{g}=0 V$, and simultaneously matching the value of the 
density obtained from $n \propto 1/R_{Hall}$ using Eq. (2) (1 component), of 
$1.6\pm0.02\times10^{12}cm^{-2}$, we find that $r\sim0.22$ and 
$q=2.5\sim2.15/0.85 = 2.53$. Finally, compute the factor: 
$qr^{2}/(qr^{2}+1)\sim1/9$. We therefore expect a reduction of a factor $\sim9$ 
for the AHE coefficient ($\sim 16,000$, yielding a lower bound that is still 
$\sim1,800$ times smaller than the p-type system\cite{2DHG}. Although the 
mobility r is on the low side in this crude estimate, it is still roughly of 
order unity. 

The difference in density and mobility of this Hall device, compared to the 
first device, could arise from inhomogeneity in the Mn atom density across the 2 
inch crystal wafer.  Although the wafer was rotated during growth, at the 
approximate speed of 1 revolution every 2 seconds, it takes less than 1 second 
to deposit a monolayer. Thus the difference in Mn atom flux across the wafer is 
accentuated.  Since Mn acts as effective acceptors, at 2/3 ML, even a 1 percent 
variation in flux could account for the observed differences.

Our results indicate that the AHE is either absent or at least very small 
compared to an equivalent system where the carrier is the hole. This seems to 
confirm the predictions made by Jungwirth et al. in \cite{JungwirthPRL2002}, 
where they show that in the case of a system with small or no spin-orbit 
interaction, the AHE coefficient should be very small compared to a system with 
large spin-orbit interaction.

As a final note, we speculate on the nature of the ferromagnetic interaction. In  
p-type carrier mediated ferromagnetic GaMnAs systems, the magnetic interaction 
is believed to be of the Zener type, where double-exchange through the hole-
impurity band gives rise to the magnetic coupling. Despite the complexity of our 
structure with both Si (n-type) and Be (p-type) dopants, all in the presence of 
Mn atoms, which act themselves as acceptors, the fact that the GaMnAs QW can be 
filled with electrons indicates that all acceptor levels are filled. Therefore, 
hole-mediation cannot take place. For the QW 2DEG in our first sample, the 
mobility and the density ($\mu\sim598cm^{2}/Vs$ and 
$n\sim0.72\times10^{12}cm^{-2}$) yield a mean free path (mfp) and Fermi 
wavelength of $l_{mfp}\sim8.30nm$ and $\lambda_{F}\sim29.5nm$, 
respectively. Since $l_{mfp} $ is significantly smaller than the $\lambda_{F}$, 
it is quite likely that conduction could occur through an impurity band as well. 
Moreover, some form of anisotropy in the exchange interaction must be in place 
to stabilize the 2-d magnetism, such as the dipolar interaction with an energy 
scale of $\sim100\mu eV$.

We would like to thank Qian Niu for useful discussions. Angelo Bove would like 
to thank Ndeye Khady Bove for her support. This work was supported by NSF DMR-
0135931.

\end{document}